\documentclass[aps,prd,eqsecnum,notitlepage,
nofootinbib,superscriptaddress]
{revtex4-2}
\global\arraycolsep=2pt
\usepackage{stmaryrd}
\usepackage{amsmath}
\usepackage{amssymb}
\usepackage{graphicx}
\usepackage{textcomp}
\usepackage{calrsfs}
\usepackage{yfonts}
\usepackage{bm}
\usepackage{xcolor}
\newcommand{\ben}{\begin{equation*}}                                           
\newcommand{\een}{\end{equation*}}
\newcommand{\bean}{\begin{eqnarray*}}                              
\newcommand{\eean}{\end{eqnarray*}}

\newcommand{\nn}{\nonumber}
\newcommand{\be}{\begin{equation}}             
                               
\newcommand{\ee}{\end{equation}}
\newcommand{\bea}{\begin{eqnarray}}                                            
\newcommand{\eea}{\end{eqnarray}}

\DeclareMathOperator{\tr}{tr}

\DeclareMathOperator{\re}{Re}
\DeclareMathOperator{\im}{Im}

\begin{document}
\title{Spontaneous Torque on an Inhomogeneous Chiral Body out of Thermal 
Equilibrium}

\author{Kimball A. Milton}
  \email{kmilton@ou.edu}
  \affiliation{H. L. Dodge Department of Physics and Astronomy,
University of Oklahoma, Norman, OK 73019, USA}

\author{Nima Pourtolami}
\email{nima.pourtolami@gmail.com}
\affiliation{National Bank of Canada, Montreal, Quebec H3B 4S9, Canada}

\author{Gerard Kennedy}\email{g.kennedy@soton.ac.uk}
\affiliation{School of Mathematical Sciences,
University of Southampton, Southampton, SO17 1BJ, UK}

\begin{abstract}
In a previous paper we showed that an inhomogeneous body in vacuum will
experience a spontaneous force if it is not in thermal equilibrium with its
environment.  This is due to the asymmetric asymptotic radiation pattern
such an object emits.  We demonstrated this self-propulsive force
 by considering an expansion in powers of the electric susceptibility: 
A torque arises in first order, but only if the material
constituting the body is nonreciprocal.
No force arises in first order. 
A force does occur for bodies made of 
ordinary (reciprocal) materials in second order.
Here we extend these considerations to the torque.
As one would expect, a spontaneous torque will also
appear on an inhomogeneous  chiral object if it 
is out of thermal equilibrium with its environment.  Once a chiral body starts
to rotate, it will experience a small quantum frictional torque, but much
more important, unless a mechanism is provided to maintain the nonequilibrium
state, is thermalization:  The body will rapidly reach thermal equilibrium
with the vacuum, and the angular acceleration will essentially become zero.
For a small, or even a large, inhomogeneous
chiral body, a terminal angular velocity will result, which seems to be
in the realm of observability.
\end{abstract}

\date\today
\maketitle

\section{Introduction}
There is extensive theoretical
literature over recent decades concerning quantum electrodynamic friction 
between two relatively moving surfaces, or between a particle or atom moving 
parallel to a conducting or dielectric surface; see 
Refs.~\cite{teod,lev,pendry,volokitin,dedkov,davies,barton,Zhao,Silveirinha,%
Pieplow15,intravaia,rev,Dedkov20,Farias,marty,
Dedkov22,Wang,Brevik22,Xu}, for
example.
But much earlier, Einstein
and Hopf \cite{EH} showed that a polarizable particle moving 
with nonrelativistic velocity $v$ through vacuum, filled with blackbody
radiation at temperature $T=\beta^{-1}$,
experiences a frictional force, which can be expressed in generalized form for
an isotropic particle with polarizability $\alpha(\omega)$:
\be
F^{\rm EH}=-\frac{v\beta}{12\pi^2}\int_0^\infty d\omega\,\omega^5
\im\alpha(\omega)\frac1{\sinh^2\beta\omega/2}.
\ee 
Here, the imaginary part of the polarizability may arise from either
intrinsic processes within the particle, or, for the case of an atom, from
fluctuations in the electromagnetic field.
In the last two decades there has been considerable work  on this effect: 
For a subset of the 
literature see Refs.~\cite{Mrkt,dedkov2,Lach,pieplow,Volokitin0,vp}.
But such a force is very small, and has never been observed.  We generalized
this result to relativistic velocities in Refs.~\cite{guoqf1,guoqf2}, where
the force might be much more appreciable.  In particular, we considered
the temperature of the particle, $T'$, to be different from that of the
background, $T$.  We found that the nonequilibrium steady state (NESS)
\cite{Reiche},  where the particle neither gains nor loses energy in
its rest frame,
requires that the ratio of these two temperatures, $T'/T$, 
has a definite value, different from  unity, 
which might prove to be a signature for quantum vacuum
friction. The NESS condition plays a role analogous to thermal equilibrium
in this dynamic context and reduces to thermal equilibrium when the relative
velocity between the body and its environment vanishes, in which case
$T'/T\to 1$. 

Even more remarkably,  a stationary
body may experience a force or torque in a nontrivial background
\cite{khandekar0,gelb}, or  
in vacuum 
\cite{krugeremig,Ott,maghrebi,pan,Khandekar,fogedby,guofan,fan,
strekha,Ge}, provided it is out of thermal equilibrium with its environment.
In particular, for a nonreciprocal body \cite{tutorial}, 
that is, one for which the electric susceptibility
has a nonsymmetric real part, a spontaneous 
torque can arise in vacuum if the temperature of the
body differs from that of the vacuum \cite{fan,strekha,torque1,GK}. No force
can arise, however, in first order in the susceptibility,
unless another body, such as a conducting plate, is 
present. Such forces and
torques owe their origin to
radiative heat transfer \cite{Zhu}.

Forces and torques on ordinary bodies made of reciprocal materials also arise,
but in higher order in the electric susceptibility.  
M\"{u}ller and Kr\"{u}ger \cite{muller} showed that a Janus ball with the
two halves made of different homogeneous reciprocal  materials would experience
a force, for which they gave an expression in the dilute approximation,
in second order in the susceptibility.
Shortly thereafter, this effect
 was confirmed  by Reid et al.\ \cite{Reid} through numerical calculations,
but those authors found a different scaling
 behavior with the radius of the ball. 
To achieve a spontaneous force, to second order in electric susceptibility,
 requires that the body
 not only be out of thermal equilibrium with its environment, but also that
it be inhomogeneous, having an electric susceptibility that varies over the
body. Of course, the distribution 
of susceptibility must break reflection invariance so there
is an axis for motion.
According to Ref.~\cite{Reid}, however, the inhomogeneity requirement is not
necessary for a torque to appear;  all that is required is that the body
be chiral, that is, have a ``handedness.''
  
In this paper we continue our systematic treatment of 
quantum-thermal vacuum torques
and forces.  In Ref.~\cite{selfprop} we started from the classical expression
for the force on a dielectric body, and expanded it out to  fourth order
in generalized susceptibilities (i.e., 
the electric susceptibility or the Green's
dyadic) and quantized the result by using the fluctuation-dissipation
theorem (FDT).  In this way we obtained expressions for the force on an
inhomogeneous object out of thermal equilibrium with its vacuum environment,
which we evaluated for a number of examples.  The results were in qualitative
agreement\footnote{There appear to be different results in the literature
for the dependence of the force on the size of the Janus ball. For example,
the dependence we found 
in Ref.~\cite{selfprop} for the second-order susceptibility
 contribution to the force
disagrees with that given in Refs.~\cite{muller,Reid}, 
which do not seem to agree with each other.}
 with previous work \cite{muller, Reid, Manjavacas}.  Of course, the fact
that the body was hotter or colder than its environment meant that it would
rapidly thermalize, but even so, the resulting terminal velocity
might be observable.

Here, we return to the torque, but consider it in second order in
the electric susceptibility.  To achieve a nonzero torque in second order, 
the body
again must be out of thermal equilibrium with the vacuum, be chiral, to break
rotational and reflection
 invariance, and be inhomogeneous.  This is in contrast to the 
nonperturbative numerical work of Ref.~\cite{Reid}, which finds a torque for a 
chiral body made of isotropic and  uniform gold.

The outline of this paper is as follows.  In Sec.~\ref{SourceTorque} we
first rederive the torque on a nonreciprocal body, which occurs in first
order in  the electric susceptibility.  We then expand the expression for
the torque to second order and find that 
a torque arises in general for a reciprocal
body provided it is chiral and inhomogeneous.  In Sec.~\ref{FieldTorque}
we rederive the general torque
expression we found in Sec.~\ref{SourceTorque} by examining the radiation-zone
electromagnetic fields through the angular momentum flux vector.
We then consider two simple examples which would exhibit a torque but not
a force (so the body might be observed for some time without  it flying
away).  In Sec.~\ref{ex1} we examine what we dub a dual Allen wrench, with 
dielectric tags perpendicularly attached  to a central metal wire.  The torque
and resulting terminal angular velocity may be enhanced if the
tags be replaced by two-dimensional flags, which we discuss in Sec.~\ref{ex2}.
Conclusions round out the paper.  Two Appendices follow,
one on the general expansion scheme in susceptibilities, 
Appendix \ref{appa}, and one
on the duality between dielectric-metal and blackbody-metal composites,
Appendix \ref{appb}.

We use natural units, $\hbar=c=\epsilon_0=\mu_0=k_B=1$.

\section{Torque: Source point of view}
\label{SourceTorque}
Classically, the torque on a stationary dielectric 
body with polarization vector $\mathbf{P}$ is
given by \cite{CE}
\be
\bm{\tau}=\int (d\mathbf{r})\frac{d\omega}{2\pi}\frac{d\nu}{2\pi}
e^{-i(\omega+\nu)t}\left[\mathbf{P(r};\omega)\times \mathbf{E(r};\nu)+
P_i(\mathbf{r};\omega)(\mathbf{r}\times\bm{\nabla})E_i(\mathbf{r};\nu)
\right].
\ee
The first term here is called the internal torque and the second the
external torque, because the latter is reflective of the force on the body.

\subsection{First-order calculation}
Let us first review how this yields, in first order in the susceptibility, a
torque if the body 
is out of  thermal equilibrium with the vacuum.
To do this we either expand $\mathbf{E}$ to first order
in $\mathbf{P}$,
\begin{subequations}
\be
E^{(1)}_i(\mathbf{r};\omega)
=\int (d\mathbf{r'}) \Gamma_{ij}(\mathbf{r-r'};\omega)P_j(\mathbf{r'};\omega),
\ee
where $\bm{\Gamma}$ is the vacuum Green's dyadic,
or we expand $\mathbf{P}$ to first order in $\mathbf{E}$,
\be
P^{(1)}_i(\mathbf{r};\omega)=\chi^{\vphantom{q}}_{ij}(\mathbf{r};\omega)
E_j(\mathbf{r};\omega),
\ee
\end{subequations}
where we assume that the electric susceptibility $\bm{\chi}$ is local in space.
These expressions give the first-order linear responses
to the fluctuating fields  
$\mathbf{P}$ and $\mathbf{E}$, respectively. See Appendix \ref{appa} for the
general scheme of the expansion.
The resulting terms, due to expanding $\mathbf{P}$ once, 
denoted $(1,0)$,
or expanding $\mathbf{E}$ once,  $(0,1)$,\footnote{Here, $(n,m)$ designates
the number of iterations in the expansion of $\mathbf{P}$ and $\mathbf{E}$,
respectively.}
 are then quantized using the  FDT 
for the independent fluctuating fields:
\begin{subequations}
\bea
\langle P_j(\mathbf{r};\omega)P_l(\mathbf{r}';\nu)\rangle&=&
2\pi\delta(\omega+\nu)\delta(\mathbf{r-r}')\,\Im 
\chi^{\vphantom q}_{jl}(\mathbf{r};\omega)\coth \frac{\beta'\omega}2,
\label{fdtp}\\
\langle E_j(\mathbf{r};\omega)E_l(\mathbf{r}';\nu)\rangle&=&
2\pi\delta(\omega+\nu)\,\Im 
\Gamma_{jl}(\mathbf{r-r'};\omega)\coth \frac{\beta\omega}2,\label{fdte}
\eea
\end{subequations}
where $T=1/\beta$ is the temperature of the blackbody
environment, and $T'=1/\beta'$ is the temperature of the body,
and symmetrized products of fields are assumed.\footnote{
Note that, at zero temperature, the fluctuations are purely quantum, while,
in the limit of infinite temperature, the fluctuations are purely classical. 
 In between, the fluctuations are 
thermal but reflect the quantum nature
of the system.  This is easily seen on reinstating conventional units,
and recognizing that the thermal factor $(e^{\hbar\omega/k_B T}-1)^{-1}$
explicitly involves both $\hbar$ and $k_B$.}
The FDT requires that
the ``imaginary part'' appearing here be the anti-Hermitian part,
\be
\Im\bm{\chi}\equiv \frac1{2i}\left(\bm{\chi}-\bm{\chi}^\dagger\right).
\ee  
(The vacuum is, of course, reciprocal, which means $\Im\bm{\Gamma}=
\im\bm{\Gamma}$, the ordinary imaginary part.)
In this order, the Green's dyadics are evaluated at coincident points,
\be 
\bm{\Gamma}(\mathbf{r-r'};\omega)\to \bm{1}\left(\frac{\omega^2}{6\pi R}
+i\frac{\omega^3}{6\pi}+O(R)\right),\quad R=|\mathbf{r-r'}|\to 0,
\ee
and its gradient is odd in $\mathbf{r-r'}$. This means that the external
torque vanishes, which is the reason there can be no spontaneous force in this
order. From the internal torque, we immediately
obtain the sum of the $(1,0)$ and $(0,1)$ contributions,
\be
\tau_i=\int(d\mathbf{r})\frac{d\omega}{2\pi}\epsilon_{ijk}\re 
\chi^{\vphantom q}_{jk}(\mathbf{r};\omega) 
\frac{\omega^3}{6\pi} \left(\coth\frac{\beta\omega}2
-\coth\frac{\beta'\omega}2\right).\label{nrtorque}
\ee
Therefore, a spontaneous vacuum torque can arise  
only for a nonreciprocal material, which possesses a 
nonsymmetric real part of the electric susceptibility.\footnote{
Since the typical way such a nonreciprocity can arise is through
an external magnetic field, calling this a vacuum effect seems to be an
oxymoron.}  This is exactly the result found in 
Refs.~\cite{torque1,fan,strekha,GK}.

\subsection{Second-order torque}
To obtain a torque on a body made of reciprocal material, we have to go to 
second order in the electric susceptibility. 
To deal with the greater complexity, let
us start with the terms coming from PP fluctuations.  Using the designation 
above,  these involve two contributions, $(2,1)$ and $(0,3)$. The former is
\bea
\tau^{(2,1)}_i&=&\int (d\mathbf{r})(d\mathbf{r'})(d\mathbf{r''})\frac{d\omega}
{2\pi}\frac{d\nu}{d\pi}\epsilon_{ijk}\bigg[\chi^{\vphantom{q}}_{jm}(\mathbf{r};
\omega)\Gamma_{mn}(\mathbf{r-r'};\omega)P_n(\mathbf{r'};\omega)
\Gamma_{kl}(\mathbf{r-r''};\nu)P_l(\mathbf{r''};\nu)\nn\\
&&\qquad\mbox{}+\chi^{\vphantom{q}}_{lm}(\mathbf{r};
\omega)\Gamma_{mn}(\mathbf{r-r'};\omega)P_n(\mathbf{r'};\omega)
r_j\nabla_k\Gamma_{lp}(\mathbf{r-r''};\nu)P_p(\mathbf{r''};\nu)\bigg].
\eea
Use of the FDT (\ref{fdtp}) yields
\bea
\tau^{(2,1)}_i&=&\int (d\mathbf{r})(d\mathbf{r'})\frac{d\omega}{2\pi}
\epsilon_{ijk}\bigg[\chi^{\vphantom{q}}_{jm}(\mathbf{r};
\omega)\Gamma_{mn}(\mathbf{r-r'};\omega)\Im\chi^{\vphantom{q}}_{nl}
(\mathbf{r'};\omega)
\Gamma_{kl}(\mathbf{r-r'};-\omega)\nn\\
&&\qquad\mbox{}+\chi^{\vphantom{q}}_{lm}(\mathbf{r};
\omega)\Gamma_{mn}(\mathbf{r-r'};\omega)\Im\chi^{\vphantom{q}}_{np}
(\mathbf{r'};\omega)
r_j\nabla_k\Gamma_{lp}(\mathbf{r-r'};-\omega)\bigg]\coth\frac{\beta'\omega}2.
\eea
A small check of this is that if in the second term, the external part,
we substitute $\mathbf{r\to r+R}$, where $\mathbf{R}$ designates the position
of the center of mass of the body, we read off the force contribution found in 
Ref.~\cite{selfprop}:
\be
F^{(2,1)}_k=\int (d\mathbf{r})(d\mathbf{r'})\frac{d\omega}{2\pi}
\chi^{\vphantom{q}}_{lm}(\mathbf{r};\mu)
\Gamma_{mn}(\mathbf{r-r'};\omega)\Im\chi^{\vphantom{q}}_{np}
(\mathbf{r'};\omega)
\nabla_k\Gamma_{lp}(\mathbf{r-r'};
-\omega)\coth\frac{\beta'\omega}2.
\ee

The formula for the torque simplifies considerably if the susceptibility is
both isotropic and homogeneous,
$\chi^{\vphantom{q}}_{lm}(\mathbf{r};\omega)=\delta_{lm}\chi(\omega)$.  Then,
\be
\tau^{(2,1)}_i=\int(d\mathbf{r})(d\mathbf{r'})\frac{d\omega}{2\pi}
\epsilon_{ijk} \im[\chi(\omega)] \chi(\omega)\left[\Gamma_{jl}(\mathbf{r-r'};
\omega)\Gamma_{kl}(\mathbf{r-r'};-\omega)+\Gamma_{ml}(\mathbf{r-r'};\omega)
r_j\nabla_k \Gamma_{ml}(\mathbf{r-r'};-\omega)\right]
\coth\frac{\beta'\omega}2.
\ee
This evidently vanishes:  the second term, the external contribution, is
zero, because it is odd under interchange of $\mathbf{r}$ and $\mathbf{r'}$
[$r_j\to \frac12(r_j+r'_j)$ to survive the Levi-Civita symbol]; and the first
term (the internal term) is zero because the vacuum Green's dyadic is entirely
constructed from the vector $\mathbf{r-r'}$.  If the body is homogeneous but
not isotropic, $\bm{\chi}(\mathbf{r};\omega)=\bm{\chi}(\omega)$, it is
similarly seen that the torque around any principal axis is zero.

If we assume isotropy only, but not homogeneity, it is still true that the
internal torque vanishes, and we are left with
\be
\tau^{(2,1)}_i=\int(d\mathbf{r})(d\mathbf{r'})\frac{d\omega}{2\pi}
\epsilon_{ijk} \im[\chi(\mathbf{r'};\omega)] \chi(\mathbf{r};\omega)
\Gamma_{ml}(\mathbf{r-r'};\omega)
r_j\nabla_k \Gamma_{ml}(\mathbf{r-r'};-\omega)
\coth\frac{\beta'\omega}2.
\ee

In the same way, we can write down the (0,3) contribution,
\bea
\tau^{(0,3)}_i&=&\int (d\mathbf{r})(d\mathbf{r'})\frac{d\omega}{2\pi}
\epsilon_{ijk}\bigg[\Im[\chi^{\vphantom{q}}_{jm}(\mathbf{r};
\omega)]\Gamma_{kl}(\mathbf{r-r'};-\omega)\chi^{\vphantom{q}}_{ln}
(\mathbf{r'};-\omega)
\Gamma_{nm}(\mathbf{r-r'};-\omega)\nn\\
&&\qquad\mbox{}+\Im[\chi^{\vphantom{q}}_{lp}(\mathbf{r};
\omega)]r_j\nabla_k\left[\Gamma_{lm}(\mathbf{r-r'};-\omega)\right]
\chi^{\vphantom{q}}_{mn}
(\mathbf{r'};-\omega)
\Gamma_{np}(\mathbf{r-r'};-\omega)\bigg]\coth\frac{\beta'\omega}2.
\eea
The second, external term, indeed reproduces the (0,3) force contribution 
found in Ref.~\cite{selfprop}.  And again it is readily seen that if
the body is isotropic, so that $\chi^{\vphantom{q}}_{ij}(\mathbf{r};\omega)=
\delta_{ij}\chi(\mathbf{r};\omega)$, the internal torque vanishes, while
if the susceptibility is homogeneous, the entire torque vanishes.  However,
if the susceptibility is isotropic but inhomogeneous, the external torque
is nonzero---as it must be since there is a force in such 
a case---resulting in the torque 
due to PP fluctuations:
\be
\tau^{(0,3)+(2,1)}_i=\tau^{\rm PP}_i=-\int (d\mathbf{r})(d\mathbf{r'})
\frac{d\omega}{2\pi}\epsilon_{ijk}r_j\nabla_k
\left[\im \Gamma_{lm}(\mathbf{r-r'};\omega)\right]
\im\Gamma_{lm}(\mathbf{r-r'};\omega)X(\mathbf{r,r'};\omega)\coth
\frac{\beta'\omega}2,
\ee
where
\be
X(\mathbf{r,r'};\omega)=\im\chi(\mathbf{r};\omega)\re\chi(\mathbf{r'};\omega)
-\re\chi(\mathbf{r};\omega)\im\chi(\mathbf{r'};\omega).
\ee

The second-order
$\rm EE$ fluctuation contribution to the torque is calculated in the same 
way, and results merely in the change of sign and the replacement of the
thermal factor $\coth\frac{\beta'\omega}2$ by $\coth\frac{\beta\omega}2$.
For a body with isotropic susceptibility there is only the external torque,
which is
\be
\tau_i=\int\frac{d\omega}{2\pi}\left(\coth\frac{\beta\omega}2
-\coth\frac{\beta'\omega}2\right)\epsilon_{ijk}
\int(d\mathbf{r})(d\mathbf{r'})X(\mathbf{r,r'};\omega)
\im\Gamma_{lm}(\mathbf{r-r'};\omega)r_j\nabla_k
\im\Gamma_{lm}(\mathbf{r-r'};\omega),\label{torque1}
\ee
which yields the corresponding force found in Ref.~\cite{selfprop}.
Again, there is no torque or force unless the body is inhomogeneous.
This is in contrast to
 the conclusions of the numerical work by
Reid et al.~\cite{Reid}, who find a torque on  a homogeneous gold body that
breaks chiral symmetry, such as a pinwheel; however, their method is
nonperturbative.
The product of Green's
dyadics sandwiching the gradient operator can be written as
\be
\frac12\bm{\nabla} \frac{2\Delta(\omega s)}{(4\pi s^3)^2}
=\frac1{(4\pi)^2}\frac{\hat{\mathbf{s}}}{s^7}\phi(v),
\ee
with $\mathbf{s}=\mathbf{r-r'}$ and $s=|\mathbf{s}|$, where, 
according to Ref.~\cite{selfprop}, 
\be
\Delta(v)=(3-2v^2+v^4)\sin^2v-v(3-v^2)\sin2v+3v^2\cos^2v,
\ee
in terms of $v=\omega s$.
The result of differentiation gives (this is what we called $v^7D(v)$ in
Ref.~\cite{selfprop})
\be
\phi(v)=-9-2v^2-v^4+(9-16v^2+3v^4)\cos2v+v(18-8v^2+v^4)\sin2v.\label{phi}
\ee
Thus, Eq.~(\ref{torque1}) can be written as 
\be
\bm{\tau}=-\int\frac{d\omega}{2\pi}\left(\coth\frac{\beta\omega}2
-\coth\frac{\beta'\omega}2\right)
\int(d\mathbf{r})(d\mathbf{r'})X(\mathbf{r,r'};\omega)
\frac{\mathbf{r\times r'}}{(4\pi)^2 s^8} \phi(v).\label{torqueg}
\ee

The examples we consider in Ref.~\cite{selfprop} consisted of heterogeneous
bodies consisting of two homogeneous parts, $A$ and $B$. 
The antisymmetric susceptibility product $X$ becomes
\be
\frac12 X(\mathbf{r,r'};\omega)\to
X_{AB}(\omega)=\im \chi^{\vphantom{q}}_A(\omega)
\re\chi^{\vphantom{q}}_B(\omega)-
\re \chi^{\vphantom{q}}_A(\omega)\im\chi^{\vphantom{q}}_B(\omega),
\ee
where the spatial  support of $\chi_A$ is the volume $A$, while the 
spatial support of 
$\chi_B$ is the non-overlapping volume $B$. 
The torque in this situation can then be written as
\begin{subequations}
\label{gentorque}
\be
\bm\tau=\frac1{2\pi^2}\int_0^\infty \frac{d\omega}{2\pi}X_{AB}(\omega)
\left(\frac1{e^{\beta\omega}-1}-\frac1{e^{\beta'\omega}-1}\right)
\mathbf{J}_{AB}(\omega),\label{gen2torque}
\ee
where the geometric factor is 
\be
\mathbf{J}_{AB}(\omega)=-\int_A(d\mathbf{r})\int_B(d\mathbf{r'})
\frac{\mathbf{r\times r'}}{|\mathbf{r-r'}|^8}\phi(\omega|\mathbf{r-r'}|).
\label{gf}
\ee
\end{subequations}
That the integral is convergent is evident from the behavior of $\phi$ for
small  argument:
\begin{subequations}
\bea
\phi(v)&\sim& -\frac49 v^8+\frac{28}{225}v^{10}+\cdots,\quad v\ll 1,
\label{lowvphi}
\\
\phi(v)&\sim&-v^4+v^5\sin 2v+3 v^4\cos 2v+\cdots, \quad v\gg1.\label{hivphi}
\eea
\end{subequations}
It is worth noting that $\phi(v)+4 v^8/9$ is strictly positive.

\section{Torques from radiation-zone fields}
\label{FieldTorque}
Because of the discrepancy with the results of Ref.~\cite{Reid}, 
it would be well to compute the torque in a
different manner.  Such is provided by the local conservation law of angular
momentum \cite{CE}, 
\be
\partial_t\bm{\mathcal{J}}+\bm{\nabla}\cdot\mathcal{K}+\mathbf{r\times f}=0.
\label{clam}
\ee
Here, the torque density is $\mathbf{r\times f}$, in terms of the force
density $\mathbf{f}$,
$\bm{\mathcal{J}}$ is the  angular momentum density of the field, in vacuum,
\begin{subequations}\be
\bm{\mathcal{J}}=\mathbf{r\times G}, \quad \mathbf{G}=\mathbf{E\times H},
\ee
and $\bm{\mathcal{K}}$ is the angular momentum flux tensor
\be \bm{\mathcal{K}}=-\mathbf{T\times r},\quad
\mathbf{T}=\frac{E^2+H^2}2\bm{1}-\mathbf{EE-HH},\label{KT}
\ee
\end{subequations}
in terms of the vacuum field momentum $\mathbf{G}$ and the electromagnetic
stress tensor $\mathbf{T}$.
For our static situation, we may ignore the time derivative term 
in Eq.~(\ref{clam}) (it
will vanish when the FDT is applied).  Now integrate this over a very large
 ball with origin at the center of the body.  If the radius $R$ of the
sphere is very large compared to distance within the body, we have
\be
\bm{\tau}=\oint d\Omega R^2\,\mathbf{\hat{R}\cdot T\times R}
=-\oint d\Omega \,R^2\,\mathbf{\hat{R}\cdot(EE+HH)\times R}.
\ee

\subsection{First-order torque}
Let us first rederive the first-order torque for a nonreciprocal body.
In that case it is clear that the $\mathbf{EE}$ contribution to the stress
tensor is the only term that survives in the radiation zone. 
(The $\mathbf{HH}$ contribution brings in extra derivatives.)  We expand the
torque as before, but now we have to go to second order in generalized
susceptibilities: $(0,2)$ and $(2,0)$ correspond to EE fluctuations, while
(1,1) corresponds to PP fluctuations.\footnote{Now $(n,m)$
refers to $n$ iterations of the first factor and $m$ iterations of the
second.}  Let's concentrate on the latter:
\be
\tau_i^{\rm PP}=-\int(d\mathbf{r})(d\mathbf{r'})\frac{d\omega}{2\pi}
\frac{d\nu}{2\pi}\hat{R}_l \Gamma_{lm}(\mathbf{R-r};\omega)P_m(\mathbf{r}
;\omega) \epsilon_{ijk}\Gamma_{jn}(\mathbf{R-r'};\nu)P_n(\mathbf{r'};
\nu)R_k e^{-i(\omega+\nu)t}.
\ee
If we apply the FDT, this turns into
\be
\tau_i^{\rm PP}=-\int(d\mathbf{r})\frac{d\omega}{2\pi}
\hat{R}_l \Gamma_{lm}(\mathbf{R-r};\omega)
\Im\chi^{\vphantom{q}}_{mn}(\mathbf{r};
\omega) \epsilon_{ijk}\Gamma_{jn}(\mathbf{R-r};-\omega)R_k
\coth\frac{\beta'\omega}2.
\ee
We need to keep the two leading terms in the Green's dyadic here (in this
case we can drop $\mathbf{r}$ in comparison to $\mathbf{R}$):
\be
\bm{\Gamma}(\mathbf{R};\omega)\sim \frac{e^{i\omega R}}{4\pi}
\left[\frac{\omega^2}R(\bm{1}-
\mathbf{\hat{R}\hat{R}})+\frac{i\omega}{R^2}(\bm{1}-3\mathbf{\hat{R}\hat{R}})
\right], \quad R\gg r,\,r'.
\ee
The reason the higher-order term must be included here is that the $\hat{R}_l$
annihilates the leading term in the first Green's dyadic. Integrating over
the large sphere gives
\be
\oint d\Omega\, R_l R_m=\frac{4\pi}3\delta_{lm}R^2,
\ee
so then all the $R$ dependence cancels out in the radiation zone, and we
are left with
\be
\tau_i^{\rm PP}=\frac1{6\pi}\int (d\mathbf{r})\frac{d\omega}{2\pi}
i\omega^3\epsilon_{ijk}\Im\chi^{\vphantom{q}}_{kj}(\mathbf{r};\omega)
\coth\frac{\beta'\omega}2.
\ee
The only part of the anti-Hermitian part of the susceptibility that survives
in this integral is the real, antisymmetric part of the susceptibility:
\be
\im\Im\chi^{\vphantom{q}}_{kj}=-\frac12\left[\re\chi^{\vphantom{q}}_{kj}-
\re\chi^{\vphantom{q}}_{jk}\right],
\ee
so, introducing the mean polarizability
\be
\alpha_{kj}(\omega)=\int (d\mathbf{r}) \chi^{\vphantom{q}}_{kj}(\mathbf{r};
\omega),
\ee
we have
\be
\tau_i=\epsilon_{ijk}\int \frac{d\omega}{2\pi}\frac{\omega^3}{6\pi}
\re\alpha_{jk}(\omega)\left(\coth\frac{\beta\omega}2-\coth\frac{\beta'\omega}2
\right),
\ee
where we have included the corresponding EE-fluctuation 
terms, $(2,0)$
and $(0,2)$.  The only additional feature those terms bring in is that, because
of the imaginary part of the Green's dyadic that arises because of the FDT,
we must recognize the asymptotic replacement
\be
e^{i\omega R}\sin\omega R\to \frac{i}2,\quad e^{-i\omega R}\cos\omega R\to
\frac12.
\ee
Thus, we have rederived, but by a rather more elaborate calculation, the
nonreciprocal result (\ref{nrtorque}).

\subsection{Second-order torque}
When we go to second order, we have to proceed more delicately.  In particular,
the $\mathbf{HH}$ part of the stress tensor contributes equally to the
radiated angular momentum, so let us start with that.  Classically, the
corresponding contribution to the torque is
\be
\tau^H_i=-\int d\Omega \int\frac{d\omega}{2\pi}\frac{d\nu}{2\pi}R^2\, 
\mathbf{\hat{R}\cdot H(r;\omega)\, (H(r;\nu)
\times R})_i\, e^{-i(\omega+\nu)t},
\quad \mathbf{H(r};\omega)=\frac1{i\omega}\bm{\nabla}\times 
\mathbf{E(r};\omega).
\ee
We expand $\mathbf{E}$, as before, in successive powers of the generalized
susceptibilities.  The $(3,1)$ and $(1,3)$ expansions refer to PP fluctuations,
which we will detail for the case of isotropic susceptibility.   The $(3,1)$
contribution is
\bea
\tau^{H(3,1)}_i&=&-\oint d\Omega\,R^2
\int(d\mathbf{r})(d\mathbf{r'})(d\mathbf{r''})\frac{d\omega}
{2\pi}\frac{d\nu}{2\pi}\hat{R}_l\frac{\epsilon_{lmn}}{i\omega}
\nabla_m\Gamma_{nk}
(\mathbf{R-r};\omega)\chi(\mathbf{r};\omega)\,\Gamma_{kp}(\mathbf{r-r'};\omega)
P_p(\mathbf{r'};\omega)\nonumber\\
&&\qquad\times\frac{\epsilon_{iqr}}{i\nu}\epsilon_{qst}
\nabla_s\Gamma_{tu}(\mathbf{R-r''};\nu)P_u(\mathbf{r''};\nu)R_r 
\,e^{-i(\omega+\nu)t}\nonumber\\
&=&\mbox{}-\oint d\Omega 
\,R^2\int(d\mathbf{r})(d\mathbf{r'})\frac{d\omega}{2\pi}
\hat{R}_l\frac{\epsilon_{lmn}}{i\omega}\nabla_m\Gamma_{nk}
(\mathbf{R-r};\omega)\chi(\mathbf{r};\omega)\,\Gamma_{kp}(\mathbf{r-r'};\omega)
\im\chi(\mathbf{r'};\omega)\nonumber\\
&&\qquad\times\frac{\epsilon_{iqr}}{-i\omega}\epsilon_{qst}
\nabla_s\Gamma_{tp}(\mathbf{R-r'};-\omega)R_r\coth\frac{\beta'\omega}2,
\label{3.14}
\eea
where, in the second line, we have used the FDT.
Now we must do a better approximation for the Green's dyadic,
in principle,  keeping 
first-order corrections in both the exponent and the power terms:
\be
\Gamma_{nk}(\mathbf{R-r};\omega)\sim\frac{1-\mathbf{r}\cdot\bm{\nabla}}{4\pi}
\left[\frac{\omega^2}{R}(\bm{1}-\mathbf{\hat{R}\hat{R}})+\frac{i\omega}{R^2}
(\bm{1}-3\mathbf{\hat{R}\hat{R}})\right]_{nk}e^{i\omega(R-
\mathbf{\hat{R}\cdot r})}.
\ee
In this case, however, neither the $\mathbf{r}\cdot \bm{\nabla}$ nor the 
subleading $i\omega/R^2$ terms need be retained, but this is not true for the
$\mathbf{EE}$ terms.
Now the effect of the first gradient in Eq.~(\ref{3.14}) 
is confined to  the exponential:
\be
\frac1{i\omega}\nabla_m e^{i\omega(R-\mathbf{\hat{R}\cdot r})}
=\left(\hat{R}_m
-\frac{\mathbf{r}}{R}\cdot(\bm{1}-\mathbf{\hat{R}\hat{R}})_m\right)
e^{i\omega(R-\mathbf{\hat{R}\cdot r})}.\label{gradonexp}
\ee
The terms involving $R_m$ vanish by virtue of the Levi-Civita symbol.
Thus, Eq.~(\ref{gradonexp}) already provides the necessary $\mathbf{r}$ 
factor.  For the third Green's dyadic,  we merely keep
the leading term:
\be
\frac{\nabla_s}{-i\omega} e^{-i\omega(R-\mathbf{r'\cdot\hat{R}})}
\sim\hat{R}_s e^{-i\omega(R-\mathbf{r'\cdot\hat{R}})}.
\ee
Putting this together, we have
\be
\tau_i^{H(3,1)}=\oint d\Omega 
\int(d\mathbf{r})(d\mathbf{r'})\frac{d\omega}{2\pi}
\frac{\omega^4}{(4\pi)^2}
\chi(\mathbf{r};\omega)\im\chi(\mathbf{r'};\omega)\coth
\frac{\beta'\omega}2e^{i\omega\mathbf{\hat{R}\cdot(r'-r)}}
\epsilon_{lmn}\hat{R}_lr_m\Gamma_{np}(\mathbf{r-r'};\omega)(\delta_{ip}-
\hat{R}_i\hat{R}_p).
\ee
Now let us again 
 abbreviate $\mathbf{v}=\omega(\mathbf{r'-r})$ and work out the
averages, 
\begin{subequations}
\bea
\oint d\Omega \hat{R}_l e^{i\mathbf{\hat{R}\cdot v}}
&=&4\pi i \gamma v_l,\\
\oint d\Omega \hat{R}_l\hat{R}_m\hat{R}_n e^{i\mathbf{\hat{R}\cdot v}}&=&
4\pi i\left[\alpha v_l v_m v_n+\beta\left(v_l\delta_{mn}+v_m\delta_{nl}+
v_n\delta_{lm}\right)\right],
\eea
\end{subequations}
The coefficients $\gamma$, $\beta$ and $\alpha$ are immediately obtained by
differentiating 
\be
\oint d\Omega \,e^{i\lambda\mathbf{\hat{R}\cdot v}}=\frac{4\pi}{v\lambda}
\sin v\lambda,
\ee
once or three times, with respect to $i\lambda$.
The relevant results are
\begin{subequations}
\bea
\gamma&=&\frac{\sin v}{v^3}-\frac{\cos v}{v^2},\\
\alpha&=&-15\frac{\sin v}{v^5}+15\frac{\cos v}{v^4}+6\frac{\sin v}{v^3}
-\frac{\cos v}{v^2},\qquad\mbox{and}\\
\beta&=&3\frac{\sin v}{v^5}-3\frac{\cos v}{v^4}-\frac{\sin v}{v^3}.
\eea
\end{subequations}
As we will immediately see, the $\alpha$ term does not contribute.
Finally, we can write the internal Green's dyadic as
\be
\bm{\Gamma}(\mathbf{r-r'};\omega)=f(v)\mathbf{\hat{v}\hat{v}}+g(v)\bm{1},
\ee
with
\begin{subequations}
\bea
f(v)&=&\frac{\omega^3}{4\pi v^3}(3-3iv-v^2)e^{iv}\qquad \mbox{and}\\
g(v)&=&-\frac{\omega^3}{4\pi v^3}(1-iv-v^2)e^{iv}.
\eea
\end{subequations}
Then, it is readily seen that this contribution to the torque is
\bea
\bm{\tau}^{H(3,1)}&=&
\frac{1}{4\pi}\int(d\mathbf{r})(d\mathbf{r'}) \frac{d\omega}
{2\pi}\omega^5\re\chi(\mathbf{r};\omega)\im\chi(\mathbf{r'};\omega)
(\mathbf{r\times r'})\im[\gamma g(v)+\beta f(v)]\coth\frac{\beta'\omega}2\nn\\
&=&\frac1{(4\pi)^2}\int(d\mathbf{r})(d\mathbf{r'})\frac{d\omega}{2\pi}
\omega^8X(\mathbf{r,r'};\omega)(\mathbf{r\times r'})\frac1{4v^8}\phi(v)
\coth\frac{\beta'\omega}2, \qquad 
\eea
where $\phi$ is the function defined in Eq.~(\ref{phi}). This
result is exactly 1/4 of the PP part of Eq.~(\ref{torqueg}).
Indeed, following precisely the same procedure detailed here, the
$H$ $(1,3)$, the $E$ $(1,3)$, and the 
$E$ $(3,1)$ contributions are all the same, giving a PP
contribution to the torque exactly as found in Eq.~(\ref{torqueg}).  We
leave it to the reader to verify that the EE fluctuations give the 
corresponding structure, with the replacement of the thermal factor
$\coth\beta'\omega/2\to-\coth\beta\omega/2$. 
Note that only the $(2,2)$ $\mathbf{EE}$ and $\mathbf{HH}$ contributions are 
nonzero.\footnote{To resolve any notational confusion, recall
that $E$ or $\mathbf{EE}$ refers to a term in the 
stress tensor (\ref{KT}),
while EE refers to the FDT contribution (\ref{fdte}).}
As in Sec.~\ref{SourceTorque}, again we conclude that
inhomogeneity is required for both torque and force, in second
order.

\section{Example 1: Dual Allen wrench}
\label{ex1}
\begin{figure}
\centering
\includegraphics[width=.5\textwidth, trim = 1cm 19cm 11cm 1cm, clip]{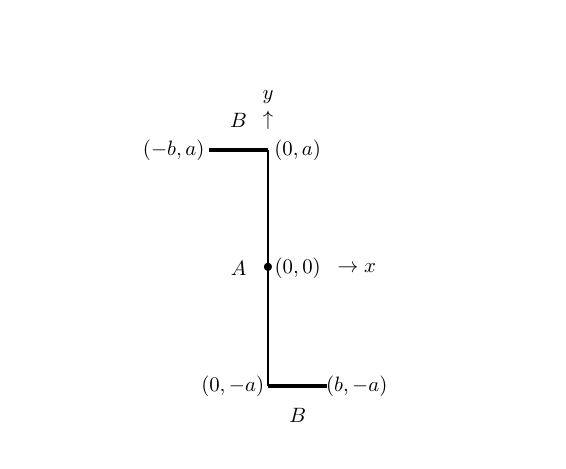}
\caption{A reflected L-shaped body 
(``dual Allen wrench'')  composed of a thin wire $A$ (cross-sectional
area $S_A$) of length $2a$ connected at the ends to perpendicular equal-length
wires $B$  (cross-sectional area $S_B$) each of length $b$.  Shown also
are the Cartesian coordinates of the center and the endpoints of the wire
segments.
The two components have respective susceptibilities $\chi^{\vphantom{q}}_A$, 
$\chi^{\vphantom{q}}_B$.  This object is evidently invariant under reflection
in the origin (0,0), but is chiral, so it will experience a quantum vacuum
torque about an axis perpendicular to the plane of the figure,
but not a force.}
\label{fig1}
\end{figure}

\begin{figure}
\includegraphics{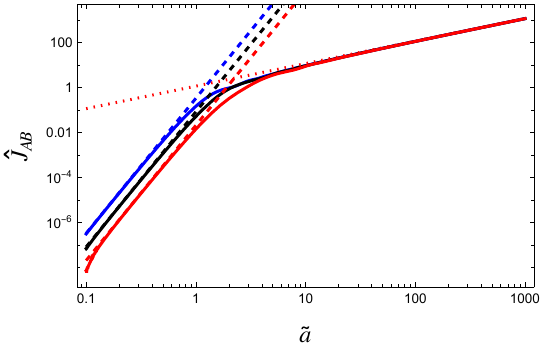}
\caption{The geometrical factor $\hat{J}_{AB}$
in Eq.~(\ref{phiint})  plotted in terms of the scaled half-length $\tilde{a}$ 
of the central 
axis of the object in Fig.~\ref{fig1}.  The solid curves show
 the cases $\tilde{a}=\tilde{b}$ (central, black), $\tilde{b}=2\tilde{a}$ 
(upper, blue), and $\tilde{a}=2\tilde{b}$ (lower,
red).  Also displayed are the asymptotic values for large $\tilde{a}$ and 
$\tilde{b}$ (dotted red line), 
which show no dependence on the aspect ratio $r=b/a$, and
the behaviors for both $\tilde{a}$ and $\tilde{b}$ small,
which do show significant dependence on $r$,  by the dashed lines. The
transition between the two asymptotic regimes occurs over a relatively  
small region around $\tilde{a}=2$.}
\label{fig:IAB}
\end{figure}

We now wish to study an example of an object that can exhibit a torque but
not a net force, so one could study the rotational effect of a small object
under a microscope.  To preclude a force on an inhomogeneous object, we can
require that it be reflection invariant about a central point.  In order that
it have a torque, the body must then be chiral, meaning that it cannot be 
transformed into any 
mirror reflection by a translation or a rotation.
A simple example of such an object is shown in Fig.~\ref{fig1}.
The object is an L-shaped figure reflected in the top of the L, which we
might call a dual Allen wrench,
 constructed of thin\footnote{Thin, because we want the thickness to be less
than the skin depth for the weak susceptibility approximation to be valid.}
 wires, with a shaft $A$ of half-length $a$ lying in the
$y$ direction, and asymmetric end tags $B$ of 
length $b$ oriented along the $x$
direction. The only significant domain of integration is
along the  wires, which have small cross-sectional areas
$S_A$ and $S_B$, respectively.
The geometric integral (\ref{gf}) then points perpendicularly to the plane of
the body (the $z$ axis), and 
 is simple if we regard the cross-sectional
radius as negligible in size:
\be 
\mathbf{J}_{AB}(\omega)=2 S_A S_B \omega^4 \hat{J}_{AB}(\omega)\bm{\hat z},
\quad
\hat{J}_{AB}(\omega)= \omega^4
\int_{-a}^a dy \int_0^b dx \, x y \frac{\phi(v)}{v^8},
\quad v=\omega\sqrt{x^2+(a+y)^2}.\label{Jaw}
\ee
Due to the positivity condition noted at the end of 
Sec.~\ref{SourceTorque}, it is evident that $J_{AB}>0$, 
which means the torque (\ref{gen2torque})
is negative, clockwise in the sense of Fig.~\ref{fig1},
if the susceptibility product $X_{AB}>0$ and $T'>T$.
Because the function $\phi$ involves significant cancellations, it is not
very practical to integrate (\ref{Jaw})  numerically. 
Instead, it is convenient to adopt polar coordinates:
\be
\tilde x\equiv \omega x=\rho\cos\theta,\quad
\tilde y\equiv\omega (y+a)=\rho \sin\theta,
\ee
where the integration is restricted to the interior of the rectangle with sides
$\tilde{a}\equiv\omega a$ and $\tilde{b}\equiv\omega b$.
The result of  straightforward integration on  $\theta$ is
\be
\hat{J}_{AB}(\omega)=-\tilde{a}\int_0^{2\tilde a}d\rho
\frac{\phi(\rho)}{\rho^6}+\int_{\tilde{b}}^{\sqrt{\tilde{b}^2+4\tilde{a}^2}}
d\rho
\frac{\phi(\rho)}{\rho^6}\left[\tilde{a}\sqrt{1-\frac{\tilde{b}^2}{\rho^2}}
+\frac{\tilde{b}^2}{2\rho}\right]+\frac12\left[\int_0^{\tilde{b}}
-\int_{2\tilde{a}}^{\sqrt{\tilde{b}^2+4\tilde{a}^2}}\right] d\rho
\frac{\phi(\rho)}{\rho^5},
\label{phiint}
\ee
which holds whatever are the relative magnitudes of $2a$ and $b$. 
Most of the $\rho$ integrals can also be carried out in closed form.
This function is always positive and,
 apart from the prefactor, is plotted in Fig.~{\ref{fig:IAB}}.
    
\begin{figure}
\includegraphics{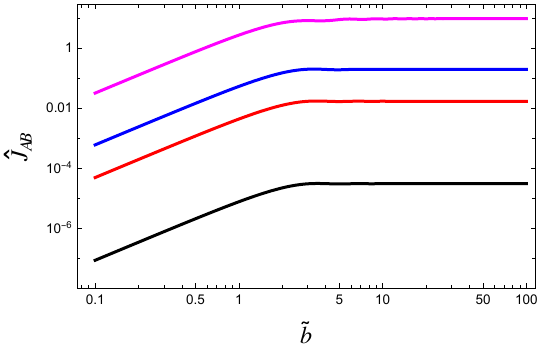}
\caption{The dependence the geometric factor $\hat{J}_{AB}$ on $\tilde{b}$
for fixed $\tilde{a}$.  The four curves are for $\tilde{a}$ equal to 0.1, 0.5,
1, and 10, from bottom to top.  
The curves are entirely similar, simply increasing with $\tilde{a}$.}
\label{fig:bdep}
\end{figure}

If both $\tilde{a}$ and $\tilde{b}$ are large, the asymptotic value  of
$\hat{J}_{AB}$  
arises only from the first integral in Eq.~(\ref{phiint}), and is
\be
\hat{J}_{AB}\sim\frac{11}{30}\pi \tilde{a},\quad \tilde{a}\gg1.
\ee 
How large an object does this correspond to?  If $T$ and $T'$ are both around
room temperature, the transition between a ``small'' and a ``large'' object
occurs when $a, b \sim 1/T\sim 10\,\mu$m, which uses the conversion
factor $\hbar c=2\times 10^{-5}$ eV cm. Perhaps unexpectedly, 
this limit is independent
of the aspect ratio $r=a/b$, and is depicted in Fig.~\ref{fig:IAB}.
The physical explanation of the large $\tilde{a}$ behavior of $\hat{J}_{AB}$
is rather simple.   The local interactions are dominated by the regions near
the corners of the object. So, holding $a$ fixed and increasing the lengths
of the tags, $b$, rather quickly saturates the integral, and the 
torque becomes independent
of $b$.  This is true even for smaller values of $\tilde{a}$, as shown in
Fig.~\ref{fig:bdep}.
The local forces at the corners also saturate as $\tilde{a}$ increases, but
the lever arm grows linearly with $a$, so the integral and torque do as well.

\begin{figure}
\includegraphics{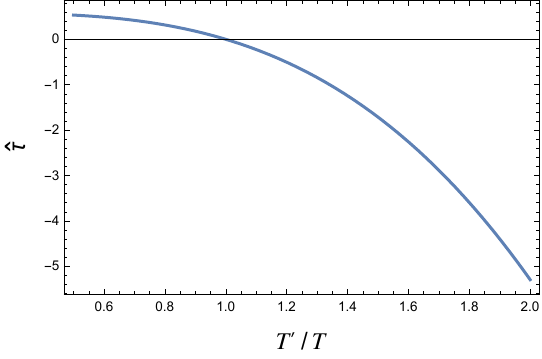}
\caption{The integral $\hat{\tau}$
 in Eq.~(\ref{hitau}) as a function of the temperature
of the body, $T'$, relative to that of a vacuum
environment at $T=300$ K, for parameters as given in the text.}
\label{hitfig}
\end{figure}

Therefore, 
for a large object of the type shown in Fig.~1, with the shaft $A$
being made of gold, described by a Drude susceptibility (nominal values taken
from Ref.~\cite{LamRey})
\be
\chi^{\vphantom{q}}_A=-\frac{\omega_p^2}{\omega^2+i\omega\nu},\quad
\omega_p=9 \,\mbox{eV},\quad \nu=0.035\,\mbox{eV},\label{Drude}
\ee
and the tags $B$ made of a dispersionless material,
we have for the torque
\be
\tau=\frac{11}{60 \pi^2} S_AS_B a \nu^4\omega_p^2
\chi^{\vphantom{q}}_B\int_0^\infty dx
\frac{x^4}{x^2+1}\left(\frac1{e^{\beta\nu x}-1}-\frac1{e^{\beta'\nu x}-1}
\right).\label{hitau}
\ee
For a wire of circular radius 50 nm,
which is approximately the skin depth of gold  \cite{selfprop}, with $a=1$ cm, 
the prefactor, $\tau_0$,  evaluates to $7 \times 10^{-22}$ N-m. 
The integral, denoted $\hat\tau$, is shown in Fig.~\ref{hitfig}.
Such a torque might well be observable. 

\begin{figure}
\includegraphics{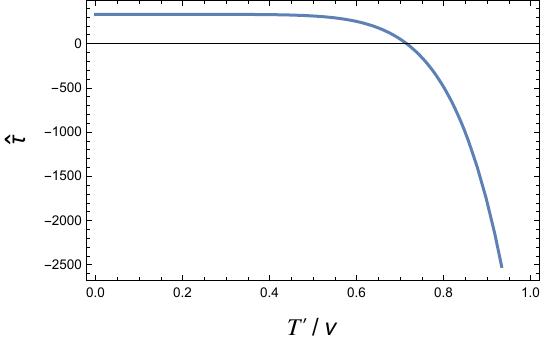}
\caption{The torque on a small Allen wrench, $\hat\tau$, 
apart from the prefactor, for the
environmental vacuum temperature of 300 K for the configuration 
considered above, with $\nu=0.035$ eV.
For $a=b=1\,\mu$m, and a common cross-sectional radius of the wires being
40 nm (dictated by the skin depth \cite{selfprop}), the prefactor in 
Eq.~(\ref{smalltorque}) 
evaluates to $3.4\times 10^{-16}$ nN-$\mu$m, which is rather large.}
\label{fig:lowtau}
\end{figure}

On the other hand, as is most easily seen from the Cartesian form of 
$J_{AB}$ (\ref{Jaw}) 
using the small $v$ behavior of $\phi$ seen in Eq.~(\ref{lowvphi}),
 the behavior of the geometric factor for a small object is
\be
\hat{J}_{AB}=\frac{56}{675}\tilde{a}^6 r^2, \quad r=b/a,
\quad \tilde{a},\,\tilde{b}\ll 1.
\ee
The contribution to $J_{AB}$ from the
 leading term in $\phi(v)$ in the small argument limit (\ref{lowvphi}) 
vanishes by symmetry, so this result reflects the $v^{10}$ term.
This limit
is also displayed in Fig.~\ref{fig:IAB}.  It agrees well with the exact
evaluation for $\tilde{a}<1$.  In this regime we can also readily calculate
the torque for the same model for the constitution of the chiral object:
\be
\tau=\frac{28}{675\pi^3}\chi^{\vphantom{q}}_B\nu^9\omega_p^2S_AS_B a^4b^2
\left[f_9(t)-f_9(t')\right],\quad
f_n(t)\equiv\int_0^\infty dx  \frac{x^n}{x^2+1}\frac1{e^{x/t}-1},  
\label{smalltorque}
\ee
where the dimensionless torque  is 
\begin{subequations}
\be \hat{\tau}=f_9(t)-f_9(t'),\quad t=\frac{T}{\nu},\quad t'=\frac{T'}{\nu},
\ee
and explicitly
\be
f_9(t)=\Gamma(8)\zeta(8)t^8-\Gamma(6)\zeta(6)t^6+\Gamma(4)\zeta(4)t^4-
\Gamma(2)\zeta(2)t^2
-\frac12\left[\psi\left(\frac1{2\pi t}\right)+\ln 2\pi t+\pi t\right].
\label{deff}
\ee
\end{subequations}
The temperature dependence of $\hat\tau$ is shown in Fig.~\ref{fig:lowtau}.

As noted below Eq.~(\ref{Jaw}),
if the body is hotter than the environment, the 
torque is always negative, that is, clockwise; 
this is because the 
underlying local forces are always toward the metal side, that is, in the $+x$ 
direction at the top, and the $-x$ direction at the bottom, using the 
coordinate system of Fig.~\ref{fig1}.

\subsection{Terminal angular velocity}
A chiral object, once set in rotation by a quantum vacuum torque, 
will feel quantum 
frictional forces opposing the motion.  More important, however, will be the
cooling of the object, if it is hotter than the environment,
or heating otherwise.  Both of these
effects will cause the object to reach a final terminal angular velocity.

\begin{figure}
\includegraphics{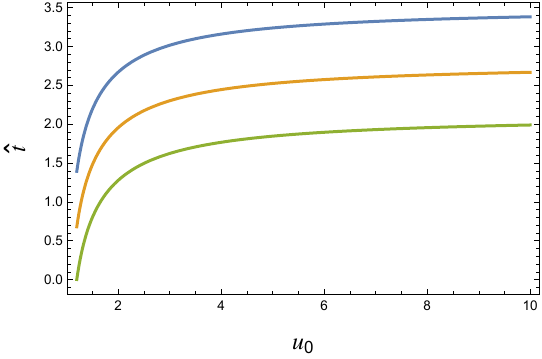}
\caption{Time taken (scaled by the prefactor $t_0$) for the object shown
in Fig.~\ref{fig1} to cool from an initial temperature $T_0'=u_0 T$, to a
final temperature $T_1'=u_1 T$.  Here $T$ is taken to be room temperature,
300 K, $\nu=0.035$ eV,
and the three curves are for different final temperature ratios
$u_1=1.05$, 1.10, and 1.20, shown from top to bottom.  The curves are similar,
and plateau for large $T'_0$.}
\label{fig:cooltime}
\end{figure}

The initial angular acceleration $\alpha$ 
about the $z$ axis is given by
\be
I\alpha =\tau,\label{reom}
\ee 
where, for the object pictured in Fig.~\ref{fig1}, the moment of inertia is
\be
I=m_A \frac13 a^2+m_B\left(a^2+\frac13 b^2\right)
=\rho_A S_A\frac23 a^3+\rho_BS_B 2b\left(a^2+\frac13 b^2\right),
\ee
with $m_A$ and $m_B$ being the total masses of the $A$ and $B$ portions of the 
object.  Let us suppose, in the perturbative spirit in which we are proceeding,
that the cooling comes only from the metal, since in our simplified 
model only the
$A$ part of the object has an imaginary part to its susceptibility.\footnote{
Realistically, this is at odds with the fact that the emissivity of a 
dielectric is much greater than that of a metal.
This point is clarified in Appendix \ref{appb}.}  According to 
Ref.~\cite{torque1} the power radiated by a Drude metal is
\be
P(T,T')=\frac1{\pi^2}2a S_A\nu^3\omega_p^2\, p(T,T'),
\quad p(T,T')=
f_3(T/\nu)-f_3(T'/\nu).\label{ptt}
\ee
Since this is the rate of energy loss by the particle, the time $t$ 
for the body to cool from temperature $T'_0$ to $T'_1$, $T'_0>T'_1>T$, is
\cite{torque1}
\be
t=\int_{T'_0}^{T'_1} dT' \frac{C_V(T')}{P(T,T')}=t_0 \hat{t}.\label{ctime}
\ee
If we suppose the specific heat
 $C_V(T)\approx 3N$, $N$ being the number of atoms in $A$,
which is the result of the Debye model for temperature 
well above the Debye
temperature, $\Theta\approx 170$ K for gold (atomic mass $m_{\rm Au}$),
\begin{subequations}
\be
t_0=\frac{3 \pi^2 \rho_{\rm Au} T}{m_{\rm Au}\omega_p^2\nu^3}
\approx 30\,\mu\mbox{s},\label{cooltime30}
\ee
putting in parameters appropriate for gold.  The remaining dimensionless
integral is
\be
\hat{t}(u_0,u_1;T)=\int_{u_0}^{u_1} du \frac1{p(u;T)},\quad u=\frac{T'}{T}.
\label{hattie}
\ee
\end{subequations}
The latter is readily integrated and is shown in Fig.~\ref{fig:cooltime}.
It actually takes an infinitely long time to cool the body to
equilibrium with the environment, but one can get arbitrarily close to the
ambient temperature in a finite time.

\begin{figure}
\includegraphics{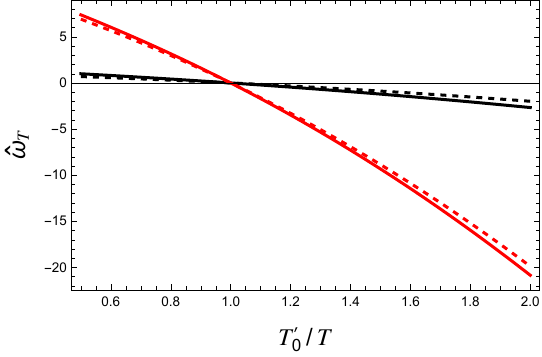}
\caption{The terminal angular velocity for the object shown in Fig.~\ref{fig1},
apart from the prefactor, (\ref{pf}), as a function of the initial temperature
of the body relative to the vacuum temperature.  The latter is taken to be
300 K in the curves with lower slope (black)  and 3000 K in the higher
slope curves (red). In both cases, $\nu$ is taken to be
$0.035$ eV.
The solid curves are the exact numerical integration, while the dashed lines
are for the high-temperature approximation.
The relative error is much smaller in the latter case.}
\label{fig:termav}
\end{figure}

Because of this cooling, the quantum torque on the body gradually decreases, 
until it finally reaches a terminal angular velocity given by integrating 
Eq.~(\ref{reom})
\be
\omega_T=\frac1I\int_0^\infty dt\,\tau(T'(t),T)=\frac{t_0\tau_0}I
\hat{\omega}_T,\quad
\hat{\omega}_T=\int_{u_0}^1
du \frac{\hat{\tau}(u;T)}{p(u;T)},\label{terminalomega}
\ee
where we have changed variable from $\hat{t}$ to $u$ according to 
Eq.~(\ref{hattie}).  Using Eq.~(\ref{hitau}) for a large object,
and ignoring the mass of the dielectric tags, we find
the prefactor in $\omega_T$ to be roughly, 
for a wire of circular cross-sectional radius 50 nm and length 1 cm,
\be
\frac{t_0 \tau_0}I\approx \frac{33}{40}\frac{S_B\nu T}{m_{\rm Au}a^2} \approx
5\times 10^{-10}\, \mbox{s}^{-1}, \label{pf}
\ee
which would seem to be undetectably small.
The integral, $\hat{\omega}_T$, is shown in Fig.~\ref{fig:termav}, and
does not change this conclusion.  Note that the rotation changes from
clockwise to counterclockwise if the 
 object is colder than the environment, because it is then absorbing
radiant energy from the environment.

\begin{figure}
\includegraphics{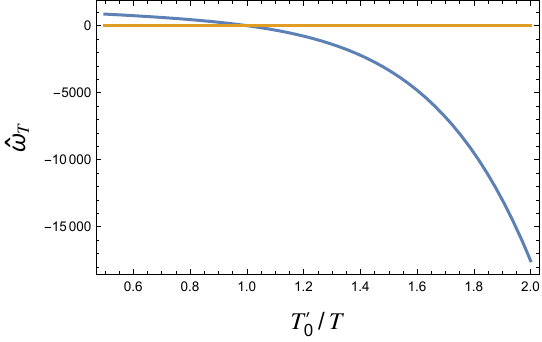}
\caption{The dimensionless integral $\hat{\omega}_T$, Eq.~(\ref{hatomegat}),
in the terminal angular velocity
for a small Allen wrench object as a function of the initial temperature $T'_0$
of the object, relative to the temperature of the vacuum background $T=300$ K,
$\nu=0.035$ eV. This integral
is typically much larger than the corresponding $\hat{\omega}_T$ 
shown in Fig.~\ref{fig:termav} for a large Allen wrench.}
\label{smtav}.  
\end{figure}

A more observable effect will arise for a small object, where the torque is
given by Eq.~(\ref{smalltorque}).  The initial acceleration when $\tilde{a}$,
$\tilde{b}\ll 1$ is, for the representative case that $S_A=S_B$, $a=b$,
\be
\alpha_0=\frac\tau{I}=\frac{14}{225\pi^3}\chi_B\nu^9\omega_p^2 S_A a^3
\frac1{\rho_A+4\rho_B}\left[f_9(T/\nu)-f_9(T'/\nu)\right]
\equiv\frac{\tau_0}I\hat{\tau},
\ee
where $f_9$ is the function defined in Eq.~(\ref{deff}).
The prefactor $\tau_0$ 
for our nominal values for gold, and $a=b=1\,\mu$m, is about
$5\times 10^{-3}$ s$^{-2}$.  The cooling time 
prefactor $t_0$ does not depend on the
dimensions of the $A$ wire, so is still 30 $\mu$s, as given in 
Eq.~(\ref{cooltime30}). 
The terminal angular velocity, given by
Eq.~(\ref{terminalomega}), where the prefactor is now $t_0\tau_0/I\approx 
2\times 10^{-7}$ s$^{-1}$, is  nearly three orders of magnitude bigger than
the corresponding prefactor for the large-argument limit (\ref{pf}).  The 
remaining integral is
\be
\hat{\omega}_T=\int_{u_0}^1 du \frac{f_9(t)-f_9(t')}{f_3(t)-f_3(t')},
\quad t=\frac{T}\nu,\quad t'=\frac{T'}\nu, \label{hatomegat}
\ee
where the functions $f_{2k+1}(t)$, with $k$ a positive integer, 
are given in Eq.~(\ref{smalltorque}), which are explicitly
\be
f_{2k+1}(t)=\Gamma(2k)\zeta(2k)t^{2k}-\Gamma(2k-2)\zeta(2k-2)t^{2k-2}+\cdots
+(-1)^{k+1}\Gamma(2)\zeta(2)t^2+
(-1)^{k+1}
\frac12\left[\psi\left(\frac1{2\pi t}\right)+\ln 2\pi t+\pi t\right].
\label{fk}
\ee
The resulting numerical integral for $\hat{\omega}_T$ is shown in 
Fig.~\ref{smtav}.  So, for the object initially at twice the ambient 
temperature, the terminal angular velocity would be $3\times 10^{-3}$ s$^{-1}$,
which should be readily observable.

\section{Example 2: Dual flag}
\label{ex2}
\begin{figure}
\includegraphics[width=.5\textwidth, trim = 1cm 19cm 11cm 1cm, clip]{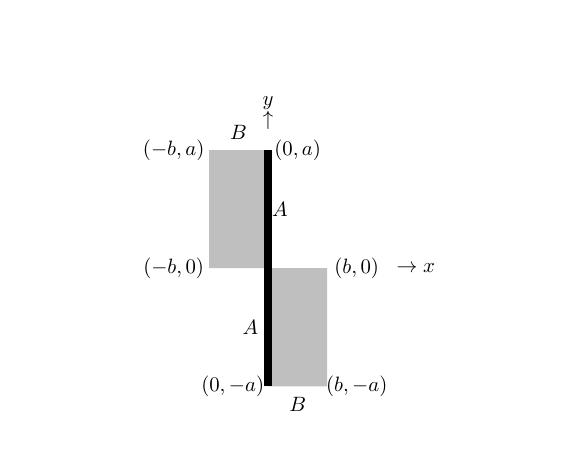}
\caption{A two-dimensional chiral object, consisting of a central metallic
wire $A$ of length $2a$ and cross-sectional area $S_A$,
 from which dielectric flags,  of height $a$, 
width $b$,  and thickness $L_B$,
extend to the left for the upper flag, and to the right for the lower flag.
The whole assembly is reflection invariant about the central point, denoted
by $(0,0)$, so there is no net self-propulsive force on the object.
}
\label{dualflag}
\end{figure}

We now consider a two-dimensional model, in which  the tags of the first
example are replaced by thin plates, 
which we will call flags, with sides $b$ and $a$, and thickness 
$L_B$, as shown in Fig.~\ref{dualflag}.

\begin{figure}
\includegraphics{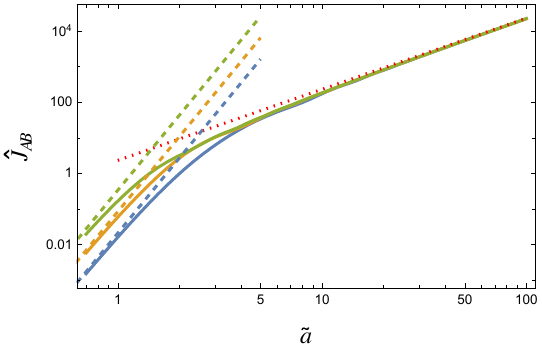}
\caption{Reduced geometric integrals $\hat{J}_{AB}$ as a function of 
$\tilde{a}=a\omega$ are shown for $b=a/2$, $b=a$, and $b=2a$ from bottom to
top.  The dotted red line shows the behavior for large $\tilde{a}$, while
the dashed lines show the small $\tilde{a}$ behavior.  The transition between
these asymptotes occurs for $\tilde{a}\sim 1$, 
that is, around $a=10\,\mu$m for frequencies corresponding
to room temperature.  As with the Allen wrench, the dependence on 
$\tilde{b}$ disappears for large $\tilde{a}$.}
\label{Jdf}
\end{figure}

The torque on this object is still given by Eq.~(\ref{gentorque}).
Evidently, only the $z$-component of the geometric integral is nonzero,
which, rather immediately, reduces to the threefold integral, because both
flags contribute equally,
\be
J_{AB}(\omega)=2S_AL_B\int_{-a}^a dy\int_{-b}^0dx'\int_0^a dy' \frac{yx'}
{[x^2+(y-y')^2]^4}\phi\left(\omega\sqrt{x^2+(y-y')^2}\right).
\ee
We adopt dimensionless variables, 
\be
\tilde x=\omega x,\quad \tilde y=\omega y, \quad\tilde y''=\omega(y'-y),
\ee
so the $y$ integral is trivial, and we are left with
\be
J_{AB}(\omega)=-\omega^3 S_AL_B\int_0^{\tilde{b}} d\tilde {x}'\left[
2\int_0^{\tilde{a}} d\tilde{y}''(\tilde{a}^2-\tilde{y}^{\prime\prime2})
+\int_0^{2\tilde a} d\tilde{y}''(\tilde{y}^{\prime\prime2}-2\tilde{a}
\tilde{y}'')\right]\frac{\tilde{x}'}{v^8}\phi(v),\quad 
v=\sqrt{\tilde{x}^{\prime 2}+\tilde{y}^{\prime\prime2}},
\ee
where $\tilde{a}=\omega a$, $\tilde b=\omega b$.  Now we adopt polar 
coordinates,
\be
\tilde{y}''=v\sin\theta, \quad \tilde{x}'=v\cos \theta,\ee
 where, with due
attention to the limits of integration, the integral over $\theta$ can be
readily carried out.  In terms of the radial integrals
\begin{subequations}
\bea
\Phi(\alpha,\beta,n)&\equiv& \int_\alpha^\beta dv\frac{\phi(v)}{v^n},\\
\Psi(\alpha,\beta,n)&\equiv& \int_\alpha^\beta dv\frac{\phi(v)}{v^n}
\sqrt{1-\frac{\alpha^2}{v^2}},
\eea
\end{subequations}
we find
\bea
J_{AB}(\omega)&=&-\omega^3S_AL_B\bigg\{-\frac13\Phi(0,\tilde{a},4)
+\frac13 \Phi(\tilde{a},2\tilde{a},4)
-\tilde{a}\Phi(0,\tilde{b},5)+\tilde{a}\Phi\left(2\tilde{a},
\sqrt{4\tilde{a}^2+\tilde{b}^2},5\right)+2\tilde{a}^2\Phi(0,\tilde{a},6)\nn\\
&&\quad\mbox{}-\tilde{a}\tilde{b}^2\Phi\left(\tilde{b},
\sqrt{4\tilde{a}^2+\tilde{b}^2},7\right)+\frac43\tilde{a}^3
\Phi\left(\tilde{a},\sqrt{\tilde{a}^2+\tilde{b}^2},7\right)
-\frac43\tilde{a}^3\Phi\left(2\tilde{a},\sqrt{4\tilde{a}^2+\tilde{b}^2},7
\right)
\nn\\
&&\quad\mbox{}+\frac23\Psi\left(\tilde{b},\sqrt{\tilde{a}^2+\tilde{b}^2},4
\right)
-\frac13\Psi\left(\tilde{b},\sqrt{4\tilde{a}^2+\tilde{b}^2},4\right)
-\frac23(3\tilde{a}^2+\tilde{b}^2)\Psi\left(\tilde{b},
\sqrt{\tilde{a}^2+\tilde{b}^2},6\right)\nn\\
&&\quad\mbox{}+\frac13\tilde{b}^2\Psi\left(\tilde{b},
\sqrt{4\tilde{a}^2+\tilde{b}^2},6\right)
\bigg\}\label{Jabdf}.\label{jphispai}
\eea
The $\Phi$ integrals can all be given in closed form in terms of sine-integral
functions, while the $\Psi$ integrals  seem to require numerics.
In Fig.~\ref{Jdf} we show the geometric integral 
$\hat{J}_{AB}(\tilde{a},\tilde{b})$, 
defined
by $J_{AB}(\omega)=\omega^3 S_A L_B\hat{J}_{AB}(\tilde{a},\tilde{b})$.

For large $\tilde a$, the leading term comes only from $\Phi(0,a,6)$
in Eq.~(\ref{jphispai}), as in the Allen wrench model, and is, as follows from
Eq.~(\ref{hivphi}),
\begin{subequations}
\be
\hat{J}_{AB}(\tilde{a},\tilde{b})
\sim \frac{11\pi}{15}\tilde{a}^2,\quad \tilde a\gg1,
\ee
independent of $\tilde{b}$, 
while the small argument expansion (\ref{lowvphi}) gives
\be
\hat{J}_{AB}(\tilde{a},\tilde{b})\sim \frac{56}{675}\tilde{a}^5 \tilde{b}^2,
\quad \tilde{a},\,\tilde{b}\ll 1.
\ee
\end{subequations}
These are also compared to the exact results in  Fig.~\ref{Jdf}.
The results are quite comparable to the geometric integral for the Allen
wrench.  In fact, comparing the full
 geometric integrals in the small- and large-frequency limits,
\be
\frac{J_{AB}^{\rm DF}}{J_{AB}^{\rm AW}}\sim \frac{a L_B}{S_B}\begin{cases}
1,&\omega a\gg1,\\
\frac12,&\omega a\ll 1.
\end{cases}\label{dfvaw}
\ee
This indicates that the geometric integrals in the two models differ only by
roughly the geometric ratio of the size of the object divided by its thickness.
Of course, the moment of inertia 
of the larger object, the dual flag, is somewhat bigger, 
\be
I^{\rm DF}=\frac13[m_A a^2+m_B(a^2+b^2)]=
\frac23 a[\rho_A S_A a^2+\rho_B L_B b(a^2+b^2)],
\ee
where $m_i$ is the mass of the $i$th component of the object.
However, typically the mass of the metal wire
 is much larger than
that of the flags, so this results in only a small correction. 
The torque for a large dual flag ($a\gg 10 \,\mu$m for room temperature) 
is thus different from that of the Allen wrench (\ref{hitau}) 
by precisely the above factor given in Eq.~(\ref{dfvaw}):
\be
\tau=\frac{11}{60\pi^2} a^2S_A L_B \chi_B\omega_p^2\nu^4
\left[f_4(T/\nu)-f_4(T'/\nu)\right].
\ee
The cooling of the object, due primarily to the metal wire,
is unchanged.  For  $a=b=1$ cm and $L_B=50$ nm, the enhancement factor is 
\be
\frac{a L_B}{S_B}\sim \frac{a}{L_B}\sim 10^5,
\ee
resulting in a terminal velocity of $\omega_T\sim 10^{-4}\,\mbox{s}^{-1}$.
For a small object, so that $\tilde{a}\sim\tilde{b}\ll 10$ $\mu$m, the 
enhancement factor is about 10, which increases the terminal velocity
to about $\omega_T\sim 3\times 10^{-2}$ s$^{-1}$
for $T=300$ K and $T'=600$ K.
   These enhancements increase
the likelihood of observing  these terminal  angular velocities.

\section{Conclusions}
In this paper we have extended the considerations of Ref.~\cite{selfprop} to
the torque on an inhomogeneous chiral body in vacuum, out of thermal
equilibrium with the blackbody radiation.  We do this by carrying out
a perturbative expansion in powers of the electric susceptibility of the body.
In first order, a torque can only arise if the susceptibility is nonreciprocal,
which usually entails an external field of some kind.  To get a true vacuum
effect on a body made of reciprocal material in second order
 requires only one additional
requirement: the body must be inhomogeneous.  A body made of uniform material
cannot experience a vacuum torque, at least through second order.\footnote{
Ongoing investigations indicate that forces and torques
do, in fact, appear in third order.}

We develop a general formalism for the torque, both by directly calculating
the torque, and looking at the angular momentum flux carried away from the body
at infinity. The general approach also allows 
for nonisotropic susceptibilities,
and arbitrary inhomogeneity.  We restrict attention, for simplicity, to
the case when all parts of the inhomogeneous body are at a common temperature.

We then consider some simple examples of bodies
composed of two contiguous parts, each of which has a uniform isotropic
electric susceptibility.  We present explicit results when one part is a 
dispersionless dielectric and the other part is a Drude metal.  We choose
configurations for which there is no net force, but only a torque. The sense
of the torque is given by that of the local forces between the two parts,
which are dominated by the immediate region of the interface.  
If the body is hotter than the vacuum, these forces
are always directed toward the metal side, due to the high reflectivity of 
the latter.  Once the body is set rotating by the quantum vacuum torque,
it will quickly reach a terminal angular velocity due to thermalization,
unless a mechanism is provided to sustain the thermal imbalance.  However,
it seems likely that configurations can be found where such a terminal 
angular velocity could be observed in the laboratory.\footnote{We speculate 
that such
forces and torques might occur with living organisms, which indeed possess
mechanisms to keep themselves out of thermal equilibrium.}

\begin{acknowledgments}This work was supported in part by the US National
Science Foundation, grant number 2008417.
 We thank Steve Fulling,
Xin Guo, and  Prachi Parashar 
for collaborative assistance.  This paper
reflects solely the authors' personal opinions and does not represent
the opinions of the authors' employers, present and past, in any
way. For the purpose of open access, the authors have applied a
CCBY public copyright license to any Author Accepted Manuscript version
arising from this submission.
\end{acknowledgments}

\appendix

\section{Susceptibility Orders in $\mathbf{E}$ and $\mathbf{P}$ Expansions}
\label{appa}

 This Appendix shows how to obtain the terms to each order in the 
generalized susceptibilities in $E$ and $P$ expansions, in a compact form that 
easily keeps track of orders and factors. The focus is on the structure of 
terms and expressions. The exposition is symbolic: no boldface, no integrals, 
delta functions implied as factors of unity, 
transpositions include coordinate variables as well as indices.

 We begin by splitting the full $E$ and $P$ fields into their corresponding 
free (or fluctuating) ($f$) and induced ($i$) parts::
\begin{subequations}
\begin{equation}
E=E^f+E^i
\end{equation}
and
\begin{equation}
P=P^f+P^i,
\end{equation}
\end{subequations}
where
\begin{subequations}
\begin{equation}
E^i=\Gamma P
\end{equation}
and
\begin{equation}
P^i=\chi E.
\end{equation}
\end{subequations}
Note that only the free (fluctuating) parts are used in the relevant FDT.
 Then
\begin{subequations}
\begin{equation}
E=E^f+\Gamma\left(P^f+\chi E\right) \quad\implies\quad E=\left(1-\Gamma\chi\right)^{-1}\left(E^f+\Gamma P^f\right)=\sum_{n=0}^{\infty}(\Gamma\chi)^n\left(E^f+\Gamma P^f\right)=\sum_{m=0}^{\infty}E^{(m)}
\end{equation}
and
\begin{equation}
P=P^f+\chi\left(E^f+\Gamma P\right) \quad\implies\quad P=\left(1-\chi\Gamma\right)^{-1}\left(P^f+\chi E^f\right)=\sum_{n=0}^{\infty}(\chi\Gamma)^n\left(P^f+\chi E^f\right)=\sum_{m=0}^{\infty}P^{(m)},
\end{equation}
\end{subequations}
where $m$ denotes the number of 
generalized susceptibility factors in these expansions. It immediately follows 
that 
\begin{subequations}
\begin{equation}
E^{(m)}=
\begin{cases}
(\Gamma\chi)^{\frac{m}{2}} E^f,  &\text{if $m$ is even},\\
(\Gamma\chi)^{\frac{m-1}{2}} \Gamma P^f, &\text{if $m$ is odd},
\end{cases}
\end{equation}
and
\begin{equation}
P^{(m)}=
\begin{cases}
(\chi\Gamma)^{\frac{m}{2}} P^f,  &\text{if $m$ is even},\\
(\chi\Gamma)^{\frac{m-1}{2}} \chi E^f, &\text{if $m$ is odd}.
\end{cases}
\end{equation}
\end{subequations}

 Products such as $\langle P\, E\rangle$ may be similarly decomposed,
 yielding
\begin{subequations}
\begin{equation}
\langle P\, E\rangle= \sum_{m=0}^{\infty}\sum_{n=0}^{\infty} 
\left\langle P^{(m)}\, E^{(n)}\right\rangle=
\sum_{\text{$m$ even}}^{\infty}\sum_{\text{$n$ odd}}^{\infty} 
\left\langle P^{(m)}\, E^{(n)}\right\rangle+
\sum_{\text{$m$ odd}}^{\infty}\sum_{\text{$n$ even}}^{\infty} 
\left\langle P^{(m)}\, E^{(n)}\right\rangle,
\end{equation}
which becomes
\begin{equation}
\langle P\, E\rangle=\sum_{\text{$m$ even}}^{\infty}
\sum_{\text{$n$ odd}}^{\infty}\tr\left\{(\Gamma\chi)^{\frac{n-1}{2}}
\Gamma P^c\left(\Gamma^T\chi^T\right)^{\!\frac{m}{2}}\right\}
+\sum_{\text{$m$ odd}}^{\infty}\sum_{\text{$n$ even}}^{\infty}
\tr\left\{(\chi\Gamma)^{\frac{m-1}{2}} \chi E^c \left(\chi^
T\Gamma^T\right)^{\!\frac{n}{2}}\right\},
\end{equation}
\end{subequations}
where $P^c\equiv \langle P^f P^f\rangle$ and $E^c\equiv \langle E^f E^f 
\rangle$ are the free (fluctuating) field correlation functions 
(symmetrization understood), to be evaluated using the relevant FDT. 
Of course, this last expression may be written in different forms.

 Where $\nabla$ operators are involved in expressions, these should be inserted
 at appropriate places in expansions such as the above.

\section{Duality between dielectric-metal and blackbody-metal}
\label{appb}
In the text we point out that the local interactions
 between the disparate portions
of the composite body are such that the effective
forces between a dispersionless
dielectric and a Drude metal are in the direction of the metal.  In 
Ref.~\cite{selfprop} 
we saw in an example that this was also true if the dielectric
were replaced by an ideal blackbody. This is a general feature: Up to a factor,
the second-order susceptibility factor is the same for the two situations.  
Suppose $\chi_A$ is given
by Eq.~(\ref{Drude}), while $\chi_B$ is a real constant.
Then, the susceptibility factor is
\be
X_{AB}(\omega)=\chi_B\frac{\omega_p^2\nu}{\omega(\omega^2+\nu^2)}.\label{dielm}
\ee
On the other hand, an ideal blackbody has a surface susceptibility of
\cite{selfprop}
\be
\tilde{\chi}_B(\omega)=\frac{i}4\frac1{\omega+i\epsilon},\quad\epsilon\to+0,
\ee
so using that for $\chi_B$ we have
\be
X_{AB}=\frac1{4\omega t}\frac{\omega_p^2}{\omega^2+\nu^2},\label{bbm}
\ee
where $t$ is the thickness of the blackbody surface.
These two susceptibility factors, (\ref{dielm}) and (\ref{bbm}), are the same
function of $\omega$, and hence yield the same torques, up to
a constant factor.  Now the blackbody is
a perfect emitter, if $T'>T$, while the Drude metal is a good reflector, 
so it seems clear that more radiation is emitted on the blackbody side
of the composite object, resulting in a force toward the metal side.
This argument seems less ambiguous than that based on the 
dielectric-metal composite.

\end{document}